\documentclass[final,5p,times,twocolumn]{elsarticle}
\usepackage{amssymb}
\usepackage[utf8]{inputenc} 
\usepackage{amsmath}
\usepackage{relsize}
\usepackage{booktabs}

\newcommand{\ud}{\mathrm{d}}

\journal{Physica A}

\begin{document}

\begin{frontmatter}

\title{Computational study of the mechanism of Bcl-2 apoptotic switch}

\author[a]{Tom\'a\v{s} Tok\'ar} 
\ead{tomastokar@gmail.com}

\author[a]{Jozef Uli\v{c}n\'y\corref{cor1}} 
\ead{jozef.ulicny@upjs.sk}

\address[a]{Department of Biophysics, University of Pavol Jozef Safarik,\\
Jesenna 5, Kosice, 040 01, Slovakia}

\cortext[cor1]{Corresponding author}

\begin{abstract}
In spite of attention devoted to molecular mechanisms of apoptosis, the details of functioning of one crucial component - the Bcl-2 apoptotic switch - are not completely understood. There are two competing mechanisms of its internal working - the indirect activation and the direct activation. In the absence of conclusive experimental data, we have used computational modeling to assess the properties of both mechanisms and their suitability to act as a biological switch. Since the two mechanisms form opposite poles of continuum of Bcl-2 molecular interaction models, we have constructed more general model including these two models as extreme cases.
By studying relationship between model parameters and steady-state response we have found optimal interaction pattern which reproduces behavior of Bcl-2 apoptotic switch. Our results show, that stimulus-response ultrasensitivity is negatively affected by  spontaneous activation of Bcl-2 effectors. We found that ultrasensitivity requires effector’s activation, mediated by another subgroup of Bcl-2 proteins - activators. We have shown that the auto-activation of monomeric effector forms provides an ultrasensitivity enhancing feedback loop. Thorough robustness analysis revealed that the interaction pattern postulated in the direct activation hypothesis is able to conserve stimulus-response switching characteristics for wide range changes of its internal parameters. The robustness of the switch against the variation of the reaction parameter is strongly reduced for the intermediate hybrid model and even more for the indirect part of the models.
Computer simulations of the more general model presented here suggest, that stimulus-response ultrasensitivity is an emergent property of the direct activation model that is unlikely to occur in the model of indirect activation. Introduction of indirect-model-specific interactions does not provide better explanation of the Bcl-2 switch functionality  compared to the direct model.
\end{abstract}


\end{frontmatter}

\section{Introduction}
\label{sec:introd}

Apoptosis, the most known form of programmed cell death, has been observed and studied for more than century~\cite{delhalle_introduction_2003}. As a physiological mechanism, important for the survival of multicellular organisms, apoptosis plays fundamental role in embryogenesis, tissue formation and homeostasis~\cite{duke_cell_1996,vaux_cell_1999}. Proliferation and apoptosis form two opposite sides of the control of optimal number of cells, so it is no surprise, that defects of apoptosis regulation are at the core of numerous diseases, including cancer, autoimmunity and neurodegenerative disorders~\cite{duke_cell_1996,delhalle_introduction_2003}. Accumulated research showed, that apoptosis is remarkably complex and tightly regulated process, starting hours before observable characteristic morphology changes and genomic destruction process. Apoptosis may be initiated by multitude of triggering events, originating from both within the cell and the extracellular signals~\cite{strasser_apoptosis_2000}. These signals are processed through multiple signaling pathways, leading eventually the cell toward internally driven progressive cellular disassembly~\cite{strasser_apoptosis_2000}.     

Major part of the known apoptotic signaling events converge to mitochondria, where they may cause rapid permeabilization of the outer mitochondrial membrane~\cite{gulbins_role_2003,elmore_apoptosis:review_2007}. Mitochondrial outer membrane permeabilization (MOMP) is considered to be crucial event in the course of apoptosis, as a discrete event which leads to the simultaneous release of the several pro-apoptotic factors, including cytochrome-c and Smac/DIABLO~\cite{kroemer_mitochondrial_1998,kroemer_mitochondrial_2003} to the cytosol. Cytosolic cytochrome-c then associates with other cytosolic pro-apoptotic factors, such as the adaptor protein Apaf-1, to form Apoptosome complex~\cite{kroemer_mitochondrial_2003,kroemer_mitochondrial_2007}. Apoptosome accelerates pro-apoptotic processes by recruiting and activating pro-forms of initiator caspases (Caspase-9), which in turn proteolytically activate executioner caspases (Caspase-3, Caspase-7)~\cite{tait_2010}. Executioner caspases are proteases with hundreds of known protein targets ~\cite{luthi_2007}, responsible for microscopically observable disassembly of cell. This mechanism is assisted by the simultaneous Smac/DIABLO inhibition of the apoptosis inhibitors IAPs, which act normally as suppressors of caspase activities. Mitochondria thus play crucial role by sensing and integrating different apoptotic signals and processing them into eventual outcome represented by MOMP event. 

Growing evidence demonstrates, that such processing is involving proteins of Bcl-2 family~\cite{kuwana_bcl-2-family_2003,adams_bcl-2_2007}, where Bcl-2 family includes both pro- and anti-apoptotic members. Structurally, Bcl-2 family of proteins can be categorized according to the number of Bcl-2 homology domains (BH) in their $\alpha$-helical regions~\cite{gross_bcl-2_1999,cory_bcl-2_2003,mayer_mitochondrial_2003}. Four BH domains (BH1-4) are characteristic for anti-apoptotic members of the family, such as Bcl-2 itself, Bcl-xL and Bcl-w~\cite{mayer_mitochondrial_2003}. Anti-apoptotic members of the family prevent MOMP event and protect cells from a wide range of cytotoxic impacts~\cite{cory_bcl-2_2003}. Pro-apoptotic members lack the fourth BH domain (BH4) and can be further divided to BH3-only proteins and multidomain proteins~\cite{cory_bcl-2_2003,adams_bcl-2_2007}. BH3-only proteins, such as Bid, Bik, Bim, Bad, Noxa and PUMA can be activated by multiple pro-apoptotic signals and cytotoxic conditions, including cytokine deprivation or DNA damage~\cite{huang_bh3-only_2000}. Multidomain Bcl-2 proteins such as Bax, Bak and Bok are also termed as ’effectors’. Once activated, multidomain Bcl-2 proteins oligomerize and form pores in mitochondrial outer membrane (MOM). Thus formed mitochondrial apoptosis-induced channels (MAC) permeabilize mitochondrial outer membrane and lead eventually to MOMP event~\cite{tsujimoto_bcl-2_2000,kuwana_bcl-2-family_2003,kinnally_tale_2007,peixoto_mac_2009}. The occurence of MOMP event is thus dependent on interplay between these three groups of Bcl-2 family.

Several experimental studies~\cite{von_ahsen_2000,rehm_single_cell_2002,madesh_2002,rehm_real_time_2003,hellwig_2008} found that while the time delay between the application of apoptotic stimulus and the onset of MOMP event is dependent on the type and/or the intensity of the apoptotic stimulus, the kinetics of MOMP event itself remains remarkably stable and MOMP occurs very rapidly on time scale of few minutes. Regulatory mechanism of such behavior was later described in terms used in electro-mechanical engineering as snap-action, variable-delay switch~\cite{albeck_quantitative_2008,sun_2010,spencer_2011}. This type of switching behavior was attributed to caspase-induced steady state bistability and extensively studied~\cite{eissing_bistability_2004,legewie_mathematical_2006,bagci_2006}. However, according to Goldstein et al., switching properties of MOMP regulation are independent of caspase related events downstream of MOMP~\cite{goldstein_2005}. 

The coordinated activity of the Bcl-2 family proteins leading to the MOMP event is commonly denoted as Bcl-2 switch~\cite{adams_bcl-2_2007,cui_two_2008}. Although the general features of the Bcl-2 switch are known, the detailed mechanism in terms of interactions between the individual members of the family remains open question. It is not generally agreed, which interactions are substantial for the induction of MOMP. It is not clear, whether the BH3-only proteins are able to activate multidomain effectors, such as Bax and Bak directly, or whether they act indirectly, through neutralization of Bcl-2 anti-apoptotic sentinels~\cite{huang_bh3-only_2000,chen_robustness_2007,cui_two_2008}. Closely related question regards the mechanism, by which the anti-apoptotic proteins inhibit the activity of effectors. Do they bind non-activated Bax/Bak or do they bind and neutralize already activated Bax/Bak?~\cite{huang_bh3-only_2000,chen_robustness_2007,cui_two_2008}.

The answers on above posed questions depend on quantitative details and in the absence of direct experimental observations, mathematical modeling and computational simulations might not only help to elucidate the outcomes of various scenaria, but also to design such experiments. Bcl-2 apoptotic switch has been modeled at various levels of details and under various scenaria. In works of Bagci et al~\cite{bagci_2006}, Albeck et al~\cite{albeck_quantitative_2008} and Harrington et al~\cite{harrington_2008} the Bcl-2 mediated regulation of MOMP is included as a part of larger, more general model of apoptosis. In these models, only direct activation of effectors is considered, while the indirect activation mechanism is absent. It should be noted, that these models, most notably the model of Albeck~\cite{albeck_quantitative_2008}, in spite of absence of indirect activation mechanism, do provide very plausible results.   

More detailed models, focused solely on Bcl-2 apoptotic switch are subject of work of Chen et al~\cite{chen_robustness_2007} and Cui et al~\cite{cui_two_2008}. In the work of Chen et al the direct and indirect models are compared  with regard to robustness of the  model toward ultrasensitivity and the direct model is found as more plausible. In the work of Cui et al., steady-state bistability is used as the evaluation criterion. The direct model is enhanced by introduction of additional auto-activation of effectors. Both direct and indirect models were found capable of bistability, although the enhanced direct model was evaluated as the most stable. The study of the bistability in the Bcl-2 apoptotic switch was later expanded by Sun et al~\cite{sun_2010}. The results of works of Cui and Chen suggest that the direct activation of effectors provides better explanation of switching properties of Bcl-2 mediated regulation of MOMP. 

However, abovementioned works did not explore other possible scenaria. There exist the possibility, that both direct and indirect mechanisms of activation of effectors may be at work simultaneously. In addition, anti-apoptotic proteins might bind effectors before their activation, while the same proteins may neutralize the already active forms. Moreover, current models also simplify the process of MAC formation to dimer~\cite{bagci_2006,harrington_2008,cui_two_2008} or tetramer~\cite{chen_robustness_2007,albeck_quantitative_2008} stage, while experimental work of Martinez-Caballero et al~\cite{martinez-caballero_assembly_2009} shows that the MAC channel comprising of at least six Bak/Bax monomer units must be formed before the proapoptotic factors are released.

In the presented work we combined all characteristic interactions of direct and indirect models into the one, hybrid model (HM). While the idea of combining direct and indirect models has been discussed previously~\cite{leber_embedded_2007,brunelle_control_2009}, it has been never examined, whether such hybrid model could provide better explanation compared to the existing models and/or whether such model gives rise to new system properties due to inclusion of new interactions. We have investigated comprehensively the importance of individual interactions between particular roles of the individual members of the Bcl-2 family in relation to desirable switching properties. In addition, we provide here more realistic model of MAC formation via oligomerization of effectors.

\subsection{Mathematical model and its biological relevance}
\label{sec:biol_relevance}

In order to reduce the overall complexity of the model, while preserving the fundamental features of the Bcl-2 family members, our model contains only typical members of the family representing whole class of the Bcl-2 proteins with similar functionalities. Anti-apoptotic family members, such as Bcl-2, Bcl-w, Bcl-xL and Mcl-1~\cite{chipuk_do_2008,mayer_mitochondrial_2003} are represented by Bcl-2. Pro-apoptotic Bcl-2 family members are grouped according to their BH domains structure. The effectors - members containing multiple BH domains, such as Bax and Bak~\cite{huang_bh3-only_2000,mayer_mitochondrial_2003,chipuk_do_2008}, are represented by Bax. BH3-only members of Bcl-2 family - Bid, Bim, Bad, Bik, Noxa, PUMA and others are represented by Act (Activators), as they are assumed to interact with effector proteins and to induce their activation.   

Bcl-2 family of proteins can be thought as biological mechanism processing multitude of input conditions and signals into a single output signal - permeabilization of the mitochondrial outer membrane, where the activation of effector molecules leads to the permeabilization itself~\cite{kroemer_mitochondrial_1998,luo_bidbcl2_1998,desagher_bid-induced_1999,ott_mitochondria_2007}. Incoming signals include (but are not restricted to) truncation of Bid by Caspase-8 and activation of other BH3-only proteins due to diverse cytotoxic conditions, such as cytokine deprivation or DNA damage~\cite{cory_bcl-2_2003,willis_apoptosis_2007}.

Input signals are represented in our model by single input stimulus E, converting activator Act into an active form Act--a ( reaction 1):
\begin{enumerate}
\item Act $\xrightarrow{E}$ Act--a
\end{enumerate}

Inhibition of the effector protein Bax by anti-apoptotic Bcl-2 proteins is essential feature of the Bcl-2 apoptotic switch. Our hybrid model contains both paths to Bax inhibition. The reaction 2 represents the binding to inactive Bax, while the reaction 3 represents the reversible neutralization of the active form Bax--a.
\begin{enumerate}
\setcounter{enumi}{1}
\item Bcl-2 + Bax $\rightleftharpoons$ Bcl-2$\sim$Bax
\item Bcl-2 + Bax--a $\rightleftharpoons$ Bcl-2$\sim$Bax--a
\end{enumerate}

Active form of Act, Act--a, also reversibly binds anti-apoptotic Bcl-2, leading to their mutual neutralization (reaction 4)~\cite{cheng_bcl-2_2001,kim_hierarchical_2006}.
\begin{enumerate}
\setcounter{enumi}{3}
\item Bcl-2 + Act--a $\rightleftharpoons$ Bcl-2$\sim$Act--a
\end{enumerate}
Act--a, in addition, can release both Bax (reaction 5) and Bax--a (reaction 6) from their complexes with Bcl-2 by competitive binding~\cite{willis_proapoptotic_2005,willis_apoptosis_2007}. Bax and Bax--a also compete to form the complex with Bcl-2 (reaction 7).
\begin{enumerate}
\setcounter{enumi}{4}
\item Bcl-2$\sim$Act--a + Bax $\rightleftharpoons$ Bcl-2$\sim$Bax + Act--a
\item Bcl-2$\sim$Act--a + Bax--a $\rightleftharpoons$ Bcl-2$\sim$Bax--a + Act--a
\item Bcl-2$\sim$Bax--a + Bax $\rightleftharpoons$ Bcl-2$\sim$Bax + Bax--a
\end{enumerate}

In the hybrid model, two alternative routes to Bax activation are included. Spontaneous activation (reaction 8) and Act--a mediated activation (reaction 9)~\cite{kuwana_bid_2002,terrones_lipidic_2004,kuwana_bh3_2005,kim_hierarchical_2006}. Reaction 10 accounts for spontaneous conformational change suppressing the activity of Bax--a.
\begin{enumerate}
\setcounter{enumi}{7}
\item Bax $\rightarrow$ Bax--a
\item Act--a + Bax $\rightarrow$ Act--a + Bax--a
\item Bax--a $\rightarrow$ Bax
\end{enumerate}

Today it is well documented, that an activation of effectors results in MAC channel formation~\cite{tait_2010}. Martinez-Caballero and colleagues showed by patch-clamping experiments on mitochondria, that MOMP is caused by the formation of MACs~\cite{martinez-caballero_assembly_2009}. The formation and growth of the MACs was characterized as step-wise sequential recruitment of the active Bax and/or Bak monomers. In accordance with these findings, we have modeled the assembly of MAC channels in the incremental growth of homo-oligomers constituted from monomeric Bax--a units (reactions 11 and 12).
\begin{enumerate}
\setcounter{enumi}{10}
\item Bax--a + Bax--a $\rightleftharpoons$ MAC$_{2}$
\item Bax--a + MAC$_{i}$ $\rightleftharpoons$ MAC$_{i+1}$
\end{enumerate}
The minimal size of oligomers, necessary for functional MAC channel was shown to be 6 monomer units per channel. The typical size of the MAC channel is found to be around 9 monomer units~\cite{antonsson_2001,valentijn_2008,martinez-caballero_assembly_2009}. According to our simulations, the biological relevance is  diminishing with growing number of monomers.  We have verified that further growth of oligomer size does not influence qualitatively the results of this work. Thus in our simulations we limited the size of the MAC channel to maximum of 20 units.   

In addition to the above reactions, all compounds are subject of continuous degradation (reaction 13). Bcl-2, Bax and Act are at the same time produced at the rate balancing their degradation flows (reactions 14-16). 
\begin{enumerate}
\setcounter{enumi}{12}
\item (All) $\rightarrow$ 
\item $\rightarrow$ Bcl-2
\item $\rightarrow$ Bax
\item $\rightarrow$ Act
\end{enumerate}

All the reactions were modeled using mass action kinetics.

\subsection{Initial concentration and reaction rates}
\label{sec:concentrations_and_rates}

Initial concentrations of particular species have been estimated in accordance with experimental data published in literature~\cite{kuwana_bcl-2-family_2003,hua_effects_2005,dlugosz_bcl-2_2006,chen_robustness_2007}. Concentrations of representatives of anti-apoptotic Bcl-2 and effector proteins (Bcl-2 and Bax, respectively) are reported in the range of hundreds of nanomols, concentrations of effectors being two times higher than  concentrations of Bcl-2~\cite{kuwana_bcl-2-family_2003,chen_robustness_2007}. Concentrations of BH3-only proteins (denoted Act in our model) are much lower, in range of 1-20 nanomols~\cite{hua_effects_2005,dlugosz_bcl-2_2006}. All concentrations are summarized in Table 1.

The binding rates of activators and anti-apoptotic Bcl-2 proteins have been estimated from corresponding dissociation constants $K_d$ published in literature~\cite{letai_distinct_2002,walensky_activation_2004} assuming reverse reaction rate $10^{-3} s^{-1}$. The remaining reaction rates (characterizing the binding of activated and non-activated Bax by anti-apoptotic Bcl-2 proteins, the direct activation of Bax and its inactivation) have been transferred from previous models~\cite{chen_robustness_2007,chen_modeling_2007,cui_two_2008}. The degradation rate of all species was set to corresponding half-life time $t_{1/2}$ of $180 min$. Production of inactive Bax, Act and Bcl-2 proteins  was modeled by zero-order reactions parametrized to balance degradation rates under initial conditions (Tables 2 and 3).

The values of input stimulus E was set from $10^{-4}$ to $10^{-1} min^{-1}$. These values correspond to outputs of Bid truncation due to catalytic activity of Caspase-8, reported to be $\sim10^{6} M^{-1}\!\cdot\!s^{-1}$~\cite{albeck_quantitative_2008}, assuming the number of molecules of active Caspase-8 within the range $10^0$ - $10^3$ molecules per cell~\cite{eissing_bistability_2004}.

\subsection{Reduction of the hybrid model to models of direct and indirect activation}
\label{sec:reduction_of_HM}

The hybrid model constructed above allows to browse continuously the space between the two hypotheses regarding the internal mechanism of the Bcl-2 apoptotic switch. By setting the reaction rates \emph{ki} and \emph{ks} to zero, the hybrid model is reduced to the model of direct activation. In turn, setting the \emph{ki1} and \emph{kc} to zero, the hybrid model reduces to the indirect one. In section~\ref{sec:robustness_analysis} we compare the hybrid model with its reduced direct model (DM) and indirect model (IM) alternatives.

\subsection{Implementation details}
\label{sec:implementation}

All models, i.e. hybrid model, as well as its reduced direct and indirect variants were expressed in the CMDL (the Chemical Model Definition Language~\cite{ramsey_dizzy:_2005}) as well as SBML (the Systems Biology Markup Language~\cite{hucka_2003}) format. The more common SBML format of the models is provided in Supplementary material of the article. All simulations were done using CMDL/SBML ODE solver (ODEtoJava-dopr54-adaptive) provided within the Dizzy - Chemical kinetics simulation software~\cite{ramsey_dizzy:_2005}. Analysis of bistability (section~\ref{sec:add_auto_activation}) was performed using Pysces~\cite{olivier_2005}.

\section{Results}
\label{sec:results}

Cellular switches, such as the Bcl-2 apoptotic switch are converting continuous incoming signals into two mutually exclusive outputs. Their role in cell signal processing is to ensure unambiguous transition between two different cellular states~\cite{eissing_steady_2007}. One of the necessary requirements for such behavior is the ultrasensitive response to input signal~\cite{angeli_multi-stability_2003,tyson_sniffers_2003,eissing_steady_2007}. Systems with ultrasensitive responses are defined as systems with response to stimulus more sensitive than that of hyperbolic system, i.e. the system described by the Michaelis-Menten equation~\cite{ferrell_trippingswitch_1996,legewie_quantitative_2005,eissing_steady_2007}. Thus, the ultrasensitivity can serve as useful criterion of suitability of different variants of interaction models between the members of Bcl-2 family, forming together apoptotic switch. To quantify the sensitivity of particular model, we used the approach first proposed in the work of Legewie et al.~\cite{legewie_quantitative_2005} (Brief summary can also be found in Appendix). In the role of reference response, we have adopted the relative amplification coefficient, normalized to Michaelis-Menten (hyperbolic) response. Models with particular parameter setup exhibiting relative amplification coefficient $n_R$ higher than unity were classified as ultrasensitive, in agreement with criterion used in~\cite{legewie_quantitative_2005}.

The switching behavior of the Bcl-2 mediated MOMP was documented by several experimental studies~\cite{von_ahsen_2000,rehm_single_cell_2002,madesh_2002,rehm_real_time_2003,hellwig_2008}. In these studies, most of the response was quantified by measurements of the kinetics of release of pro-apoptotic factors from mitochondrial intermembrane space and/or the measurements of the kinetics of change of mitochondrial transmembrane potential. In recent modeling works~\cite{chen_robustness_2007,cui_two_2008}, the response of the Bcl-2 apoptotic switch is monitored by the measure of relative activity of Bax/Bak. In the actual work, we have adopted different measure, namely the MAC impact on the permeability of outer mitochondrial membrane. In particular, we have quantified the response as the sum of contributions of individual MAC channels comprising of more than 6 monomeric units, with contribution weights proportional to their degree of oligomerization(equation 1). The number 6 reflects the minimum number of monomers, required for the functional MAC channel, while the degree of oligomerization accounts for the experimentally observed increase of the permeability with the growing MAC channel size~\cite{martinez-caballero_assembly_2009}.  
\begin{displaymath} \tag{1}
Response = \sum_{i = 6}^{20} i \cdot MAC_{i}
\end{displaymath}
\label{Response definition}

Stimulus-response curve obtained for the full hybrid model and steady-state with reference parameters (as listed in Table 3) are plotted in Fig. 2. The calculated values of relative amplification coefficients $n_{R}\sim10^{-2}$ prove that under default parameters, the hybrid model is strongly subsensitive.

\subsection{Variations of pivotal parameters and their effects on sensitivity}
\label{sec:param_variation}

In order to investigate the robustness of the behavior of the hybrid model model of Bcl-2 switch, we have evaluated relative amplification coefficients as a function of variation of individual reaction parameters \emph{ki}, \emph{ki1}, \emph{kin}, \emph{kc}, \emph{ks} and \emph{ko} from the reference values listed in Table 3. The results (Fig. 3) demonstrate, how dramatically can variation of single reaction parameter affect the sensitivity of the model. The decrease of the parameter \emph{ks} (the spontaneous activation of Bax) 100 times against the initial value changes the relative amplification coefficient by two orders of magnitude, turning subsensitive hybrid model into ultrasensitive one. Less pronounced is the influence of the parameter \emph{kc} (characterizing the activation of Bax by Act--a) and \emph{kin} (the spontaneous inactivation of Act--a). The dependence of sensitivity to parameter variation is monotonic for most of the parameters. One notable exception is the non-monotonic dependence of sensitivity on reaction parameter \emph{ki} (the binding of Bax by Bcl-2), for which the local maximum was found at $ki\sim12.0\times10^{-6}$.

Although single parameter variation may reveal how system properties depend on the value of chosen parameter in the vicinity of an established model, it still provides limited view on the properties of the model as a whole, due to interdependence of multiple parameters. To get better understanding of the impact of individual Bcl-2 protein interactions on the switching behavior, more thorough analysis is needed. We have generated 2000 sets of randomly changed values of the reaction rates defining the direct and/or indirect submodels, namely \emph{ki}, \emph{ki1}, \emph{kc} and \emph{ks}, while keeping the rest of the parameters at their reference values. For each of such individual combination of parameters, the relative amplification coefficient $n_R$ of the full hybrid model was evaluated. The relative amplification coefficient $n_R$ for a given parameter, averaged over the variations of the remaining three parameters is depicted in Fig. 4.

As can be seen from the Fig. 4., the decrease of the parameter \emph{ks} from the reference value yields on average increased relative amplification coefficient. In case of the parameter \emph{kc}, the highest average amplification coefficient ($n_R$ average almost 0.8) was obtained for the parameter sets, in which \emph{kc} varies between 17.8 and 31.6 of the initial reference value. No such apparent dependency was observed in the case of parameters \emph{ki} and \emph{ki1}, where the mean relative amplification coefficient remains approximately the same, regardless of the variation of these parameters (data not shown).

Our results indicate that the spontaneous activation of effectors (reflected in the parameter \emph{ks}) adversely influences the sensitivity of the hybrid model. On the other hand, activation of the effectors by activators (parameter \emph{kc}) seems to improve sensitivity and thus switching behavior of the model. The findings are unfavorable for the model of indirect activation, which relies on the spontaneous activation of the effectors without an activation through activators.

Taking into account the above findings, we have adjusted the \emph{ks} and  \emph{kc} parameters for the further work to $10^{-2}$ (\emph{ks}) and $2.0\times10^1$(\emph{kc}) of their initial values. In the next step, we have investigated the influence of the details of effector inhibition on the switching properties. The parameters \emph{ki} and \emph{ki1} characterize the inhibition of the effectors before and after the effector activation, respectively. The values of relative amplification coefficient as a function of simultaneous changes in both \emph{ki} and \emph{ki1} are depicted on Fig. 5. The sensitivity of the hybrid model is only weakly influenced by the changes of \emph{ki} and \emph{ki1} within two orders of magnitude. The highest relative amplification coefficient is achieved when the value of parameter \emph{ki} is reduced, while the parameter \emph{ki1} is increased, indicating the importance of the neutralization of active effectors by anti-apoptotic proteins. The lowest relative amplification coefficient was found when both parameters were reduced simultaneously.

\subsection{Addition of the auto-activation of Bax}
\label{sec:add_auto_activation}

The experimental works of Ruffolo et al.~\cite{ruffolo_bcl-2_2003} and Tan et al.~\cite{tan_auto-activation_2006} provided evidence, that activated effectors Bak, as well as Bax can positively influence the activity of non-activated Bak and Bax, respectively. Such Bax/Bak auto-activation can enhance the activity of effectors, initially activated either spontaneously (as proposed in the hypothesis of the indirect activation) and/or by BH3-only activators (as is proposed in the direct model). Such auto-activation forms positive feedback loop, which can dramatically change the stimulus-response properties of the model. The inclusion of auto-activation mechanism makes search for the ultrasensitive setup more difficult, therefore we have refrained to include this interaction in the initial stages of the model construction. On the other hand, the addition of such positive feedback mechanism may increase the ultrasensitivity of the mechanism, or even lead to the emergence of steady-state bistability~\cite{tyson_sniffers_2003}.

At the recent state of knowledge, it is not known, whether the auto-activation of Bax type effectors is mediated by the monomeric or by oligomeric forms~\cite{leber_embedded_2007}. In existing models an auto-activation mechanism was accounted mostly in the form of an attachment of the inactive effectors into already oligomerized pores~\cite{cui_two_2008,dussmann_single-cell_2009}. In the presented work we decided to investigate both possibilities. The auto-activation mediated by monomeric effectors is modeled by the reaction auto1:
\begin{itemize}
\item[] auto1: Bax + Bax--a $\rightarrow$ Bax--a + Bax--a
\end{itemize}
while the auto-activation mediated through oligomerization was modeled similarly to previous models~\cite{cui_two_2008,dussmann_single-cell_2009} as (reaction auto2):
\begin{itemize}
\item[] auto2: Bax + MAC$_{i}$ $\rightarrow$ MAC$_{i+1}$
\end{itemize}

The auto-activation of Bax with kinetics described by the reaction rate \emph{ka} (reaction auto1) was added into the hybrid model and the dependence of the relative amplification coefficient on the value of \emph{ka} has been examined. With increasing \emph{ka} value, the relative amplification coefficient rises, followed by steep decline after the \emph{ka} passes through certain threshold (see Fig. 6).

The introduction of positive feedback into an ultrasensitive system and the consequences of such addition to steady-state properties are thoroughly discussed in numerous works~\cite{ferrell_responses_1997,ferrell_regulated_1998,angeli_detection_2004}. The emergence of bistability appears for the strength of feedback exceeding some threshold value. We found similar behavior also in the considered hybrid model. Increasing the value of \emph{ka} amplifies the strength of the positive feedback, compressing the intermediate part of the sigmoid response curve and making it steeper and higher. As \emph{ka} exceeds certain value, the bistability emerges and the model turns into a ''toggle" switch (see Fig. 7)~\cite{tyson_sniffers_2003}. Further increase of the \emph{ka} value shifts the bistable range out of the reasonable range of input stimuli. For excessive values of \emph{ka}, even basal levels of incoming stimuli cause spontaneous permeabilization of the mitochondrial membrane, which manifests in the Fig. 6 as the rapid drop of the relative amplification coefficient.

In the next step, we have replaced the reaction auto1 by the reaction auto2. We found the relative amplification coefficient only slightly influenced by the variation of \emph{ka}. This indicates, that auto-activation mediated by oligomerized effectors has only marginal effect on the sensitivity of the hybrid model.

\subsection{Robustness analysis}
\label{sec:robustness_analysis}

All essential biological mechanisms should exhibit considerably robust functionality against parameter changes~\cite{kitano_systems_2002}, so that the effects of environmental differences, polymorphism and/or mutations can be compensated~\cite{bluthgen_robust_2003}. Biological switches, amongst them the Bcl-2 apoptotic switch should preserve the ultrasensitivity as their pivotal feature in spite of variation of parameters.

To examine the robustness of the hybrid model, in the first step we have adjusted the parameter setup in accordance with the previous results, in order to obtain the desirable level of sensitivity. The adjusted parameters are summarized in Table 4. To test the robustness of the model, we have measured the relative amplification coefficient for multiple sets of reaction rates (\emph{ki}, \emph{ki1}, \emph{ki2}, \emph{kc}, \emph{ks}, \emph{kin}, \emph{ko}) and initial concentrations (Act, Bax,Bcl-2).

Variation of parameters followed the methodology of Barkai and Leibler~\cite{barkai_robustness_1997} used also in some later works~\cite{bluthgen_robust_2003,chen_robustness_2007}. The modified set of parameters was created from the reference set by multiplying each of the reference parameters by $10^{q}$, where $q$ is a random number taken from the Gaussian distribution (mean = 0.0, variance = 1.0). Relative amplification coefficient of each such set was then plotted as a function of total parameter variation $T$, defined as the total order of magnitude of parameter variation~\cite{barkai_robustness_1997,bluthgen_robust_2003,chen_robustness_2007}
\begin{displaymath} \tag{2}
T = \sum_{i = 1}^{n_{p}} \Bigg\vert \log_{10} \frac{p_{i}}{p_{i,ref}} \Bigg\vert
\end{displaymath}
where $n_P$ is the number of varied parameters, $p_{i}$ is the actual varied parameter and $p_{i,ref}$ is the corresponding reference parameter before variation. Robustness was quantified by the ratio of number of ultrasensitive responses to the total number of tried parameter sets. Similar approach was used in the work of Chen et al~\cite{chen_robustness_2007}.

Such quantification of the robustness of the model in absolute terms bears little meaning when standing alone, the robustness of the model needs to be compared with an alternative model and/or reference (provided such reference model exists). It needs to be remarked, that our model differs considerably from the models of Chen et al., therefore the robustness of these models cannot be directly compared. For our purposes, it is meaningful to compare the robustness of the full hybrid model with the reduced ones, describing the direct and indirect activation. The details of the model reduction can be found in the section~\ref{sec:reduction_of_HM}.

Around 35\% of the parameter sets (out of 3000 trials) yield ultrasensitive response (see the scatter plot in Fig. 8). Most of the utrasensitive responses (around 90\%) occurred within the moderate relative amplification coefficient (the values below 2.5). As we reduced the hybrid model set into the indirect one, none of the thousand trials yielded an ultrasensitive response. On the other hand, as we turned the hybrid model into the direct one, the fraction of sigmoidal response curves with ultrasensitive behavior raised to over 66\%. Such behavior shows that omitting the spontaneous activation of effectors improves the sensitivity of the response and points to the beneficial effect of inhibition of effector proenzymes.

While the ultrasensitive character of response against the variations of the internal switch parameters is of direct importance itself, the biological functionality of the switch requires probably also that the shape and the switching point of the response curve (determining the thresholds and/or the timing of apoptotic events) should be preserved too. Since the spontaneous activation of effectors and/or inhibition of their proenzyme forms may influence this important property, we have also investigated the influence of the variations on the position of inflection point.

We took those parameter sets, which yielded relative amplification coefficients higher than one, and identified their inflection points as those points for which the slope of the response curve reaches maximum. The statistical distribution of the inflection point coordinates shows wider spread of the values, compared to the direct model. This indicates the direct model preserves better the overall (quantitative) characteristics of the switch (Fig. 9). For both models, the inflection point is observed at the vicinity of 10\% to 20\% of the maximum (saturated) value of the response. Taken together, we found that the direct model is the one better preserving not only the ultrasensitive behavior, but also the quantitative characteristics of the switching threshold behavior.

\section{Discussion}
\label{sec:discussions}

\subsection{Steady-state ultrasensitivity originates from Bax/Bak activation by BH3-only activators, not from spontaneous activation}

The aim of our modeling/simulation was to identify those interactions between the Bcl-2 groups, which are responsible for the emerging switching behavior. Single-parameter variations, as well as simultaneous variations of multiple reaction rates revealed interesting relations between the parameter setup and the steady-state sensitivity of the hybrid model.

Based on the simulation results, we can conclude that spontaneous activation of effectors Bax/Bak play minor if any role in the Bcl-2 switch. The neutralization of the already activated effectors contributes to the switching behavior to much bigger extent than the inhibition of their proenzymes. Taken together, the direct activation mechanism gives rise to ultrasensitive response, while the indirect one doesn't.

\subsection{Monomers mediated auto-activation of Bax/Bak can strongly enhance ultrasensitivity of the Bcl-2 apoptotic switch and may give rise of bistability}

As reported in the work of Tan et al.~\cite{tan_auto-activation_2006}, already activated Bax may activate its non-active form. Similar finding was reported for Bak in work of Ruffolo et al.~\cite{ruffolo_bcl-2_2003}. The mechanism of this auto-activation is still not fully understood, the auto-activation might be mediated by monomers and/or by oligomerized Bax/Bak. Such auto-activation of effectors is sometimes referred as responsible for switching properties of Bcl-2 apoptotic switch, although the relevance of such auto-activation has been never examined. 

Autocatalytic activation of Bax/Bak effectors forms positive feedback loop, in principle capable to increase the sensitivity or even lead to bistable behavior~\cite{ferrell_regulated_1998}. To model the influence of such auto-activation on the switching properties, we have added the positive feedback loop for both mechanisms and measured the sensitivity of the switch as a function of variation of their reaction rates.

The auto-activation mediated by oligomers has no significant effect on the ultrasensitivity of the modeled system. In case of monomers mediated auto-activation, within certain values of reaction rate we observed rapid increase of sensitivity and emergence of the bistability. Further increase of the \emph{ka} values led to sudden fall of the sensitivity. This is due to the withdrawal of the bistable range of the stimulus-response curve from the reasonable range of input stimuli. Overly strong auto-activation may easily cause massive activation of effectors and - from biological point of view undesirable- high MOM permeability triggered even by basal stimuli.  

On the other side, the introduction of bistability through monomeric auto-activation means that even the hybrid model is able to manifest so-called toggle switch behavior~\cite{tyson_sniffers_2003}. In toggle switch model, the steady state response switches from one state to another as the stimulus exceeds certain point. Since there is no intermediate response between these two states, bistable toggle switch differs from the ”pushbottom switch” model which includes intermediate response (sigmoid stimulus-response curves). Although bistability is not necessary for switching behavior, according to Spencer and Sorger~\cite{spencer_2011} it is the most plausible framework for modeling of switching between two states. Hybrid model with an auto-catalytic activation mediated by monomeric effectors suits to this framework.

\subsection{The direct model is more robust, compared to the hybrid or indirect ones}

The most comprehensive model was obtained upon addition of the monomeric form of auto-activation into the hybrid model and adjustment of reaction rates to obtain strongly ultrasensitive response. The resulting hybrid model was then analyzed for the robustness of switching behavior against variability of reaction rates and initial conditions. Only 35\% of trial sets describing the full hybrid model preserved the desirable ultrasensitive response, while the restricted direct mechanism model was found much more robust, with 66\% of trial sets yielding the right switching behavior. The hybrid model failed to provide ultrasensitive behavior completely. The results of such perturbation treatment indicates, that the spontaneous activation of the effectors and their preventive inhibition has detrimental effect on the ability to keep the switching behavior under naturally occurring perturbations.

An additional support for the direct model comes from the analysis of the ability to preserve the stimulus-response dependence also quantitatively. We have hypothesized that the spontaneous autoactivation and the inhibition of the effector proenzymes might increase the model’s ability to conserve the stimulus-response dependence. The distribution of the threshold stimuli (characterized through inflection points on stimulus-response curves) showed remarkably well preserved thresholds for the direct model, while the hybrid model thresholds were more susceptible to parameter variations, disproving the hypothesis.

\subsection{Indirect model interaction may provide some other system properties, but not steady-state ultrasensitivity}

Taken together, all results of this work point to following conclusions: interaction pattern as described by the hypothesis of indirect activation is very unlikely to give rise to the switching behavior expected from the Bcl-2 apoptotic switch. On the other hand, its alternative - the direct model - provides much more plausible explanation of the experimentally observed functionality. The hybrid model, containing interaction patterns from both models is less robust with respect to ultrasensitivity, and seems to be less efficient provider of the switching function.

Despite these conclusions related to steady-state ultrasensitivity, we cannot completely rule out the indirect model interactions. The requirements of ultrasensitivty suggest that indirect model interactions, if present, must proceed at very low rates, but these interactions still may exist and provide other important system properties. Low rate spontaneous activation of effectors can make response to changing stimuli more abrupt, while still preserving desirable steady-state ultrasensitivity. This vulnerability to signaling noise at certain level of affinity can be ameliorated perhaps by continuous binding and inhibition of inactive effectors by anti-apoptotic proteins. These considerations, however, need further examination. Therefore, the study of transient system properties of the Bcl-2 apoptotic switch is the aim of our further modeling work.

\section{Acknowledgments}

This work was financially supported by grants VEGA-1/4019/07; APVV- 0449-07 and VVGS PF 12/2009/F. This work could not be done without scholarships granted to Tomas Tokar from National Scholarship Programme of the Slovak Republic and ”Hlavicka” scholarship from Slovensky plynarensky priemysel a.s., whom we wish to thank for the support. Tomas Tokar appreciate many valuable discussions with colleagues from Institute for Systems Theory and Automatic Control, University of Stuttgart, Germany.

\section{Appendix}

In metabolic control analysis, the response coefficient is defined as steady state fractional change of the system response $\Delta\!X/ X$ divided by fractional change of stimulus $\Delta\!S/S$ in limit as $\Delta\!S$ tends to zero~\cite{fell_metabolic_1992,kholodenko_quantification_1997}:
\begin{displaymath} \tag{A.1}
R^{X}_{S} = \lim_{\Delta\!S \rightarrow 0} \frac{\Delta\!X/ X}{\Delta\!S/S} = \frac{\ud { ln X}}{\ud { ln S}}
\end{displaymath}
The global sensitivity is measured through relative amplification coefficient using definition of Legewie et al.~\cite{legewie_quantitative_2005} as:
\begin{displaymath} \tag{A.2}
n_{R} = \frac{\Bigg\vert\mathlarger\int\limits^{f_{H}}_{f_{L}} {R^{X}_{S}\,\ud\!{f}}\Bigg\vert}{\Bigg\vert\mathlarger\int\limits^{f_{H}}_{f_{L}} R^{X_{r}}_{S_{r}}\,\ud\!{f}\Bigg\vert}
\end{displaymath}
Where $f$ is activated fraction defined as:
\begin{displaymath} \tag{A.3}
f = \frac{X - X_{min}}{X_{max} - X_{min}}
\end{displaymath}
$f_{H}$ and $f_{L}$ are margins of the activated fraction range and $R^{X_{r}}_{S_{r}}$ is the response coefficient of the reference response $X_{r}$, which can be any monotonically increasing or decreasing function~\cite{legewie_quantitative_2005}.

\bibliography{Computational_study_of_the_mechanism_of_Bcl-2_apoptotic_switch}

\begin{thebibliography}{10}
\expandafter\ifx\csname url\endcsname\relax
  \def\url#1{\texttt{#1}}\fi
\expandafter\ifx\csname urlprefix\endcsname\relax\def\urlprefix{URL }\fi
\expandafter\ifx\csname href\endcsname\relax
  \def\href#1#2{#2} \def\path#1{#1}\fi

\bibitem{delhalle_introduction_2003}
S.~Delhalle, A.~Duvoix, M.~Schnekenburger, F.~Morceau, M.~Dicato, M.~Diederich,
  An introduction to the molecular mechanisms of apoptosis, Annals of the New
  York Academy of Sciences 1010 (2003) 1--8.

\bibitem{duke_cell_1996}
R.~C. Duke, D.~M. Ojcius, Y.~J, Cell suicide in health and disease, Scientific
  American 275 (1996) 80--87.

\bibitem{vaux_cell_1999}
D.~L. Vaux, S.~J. Korsmeyer, T.~E. Allsopp, J.~K. Fazakerley, Cell death and
  development, Cell 96 (1999) 245--254.

\bibitem{strasser_apoptosis_2000}
A.~Strasser, L.~O'Connor, V.~M. Dixit, Apoptosis signaling, Annual Review of
  Biochemistry 69 (2000) 217--245.

\bibitem{gulbins_role_2003}
E.~Gulbins, S.~Dreschers, J.~Bock, Role of mitochondria in apoptosis,
  Experimental Physiology 88 (2003) 85--90.

\bibitem{elmore_apoptosis:review_2007}
S.~Elmore, Apoptosis: a review of programmed cell death, Toxicologic pathology
  35 (2007) 495--516.

\bibitem{kroemer_mitochondrial_1998}
G.~Kroemer, B.~Dallaporta, M.~Resche-Rigon, The mitochondrial death/life
  regulator in apoptosis and necrosis, Annual Reviews in Physiology 60 (1998)
  619--642.

\bibitem{kroemer_mitochondrial_2003}
G.~Kroemer, Mitochondrial control of apoptosis: an introduction, Biochemical
  and Biophysical Research Communications 304 (2003) 433--435.

\bibitem{kroemer_mitochondrial_2007}
G.~Kroemer, L.~Galuzzi, C.~Brenner, Mitochondrial membrane permeabilization in
  cell death, Physiological reviews 87 (2007) 99--163.

\bibitem{tait_2010}
S.~W.~G. Tait, D.~R. Green, Mitochondria and cell death: outer membrane
  permeabilization and beyond, Nature Reviews Molecular Cell Biology 11 (2010)
  621--632.

\bibitem{luthi_2007}
A.~U. Luthi, M.~S. J, The casbah: a searchable database of caspase substrates,
  Cell Death and Differentiation 14 (2007) 641--650.

\bibitem{kuwana_bcl-2-family_2003}
T.~Kuwana, D.~D. Newmeyer, Bcl-2-family proteins and the role of mitochondria
  in apoptosis, Current opinion in cell biology 15 (2003) 691--699.

\bibitem{adams_bcl-2_2007}
J.~M. Adams, S.~Cory, The bcl-2 apoptotic switch in cancer development and
  therapy, Oncogene 9 (2007) 1324--1337.

\bibitem{gross_bcl-2_1999}
A.~Gross, J.~M. McDonnell, S.~J. Korsmeyer, Bcl-2 family members and the
  mitochondria in apoptosis, Genes \& Development 13 (1999) 1899--1011.

\bibitem{cory_bcl-2_2003}
S.~Cory, D.~C.~S. Huang, J.~M. Adams, The bcl-2 family: roles in cell survival
  and oncogenesis, Oncogene 22 (2003) 8590--8607.

\bibitem{mayer_mitochondrial_2003}
B.~Mayer, R.~Oberbauer, Mitochondrial regulation of apoptosis, News in
  Physiological Sciences 18 (2003) 89--94.

\bibitem{huang_bh3-only_2000}
D.~C.~S. Huang, A.~Strasser, Bh3-only proteins - essential initiators of
  apoptotic cell death, Cell 103 (2000) 839--842.

\bibitem{tsujimoto_bcl-2_2000}
Y.~Tsujimoto, S.~Shimizu, Bcl-2 family: Life-or-death switch, FEBS Letters 466
  (2000) 6--10.

\bibitem{kinnally_tale_2007}
K.~Kinnally, A.~B, A tale of two mitochondrial channels, mac and ptp, in
  apoptosis, Apoptosis 12 (2007) 857--868.

\bibitem{peixoto_mac_2009}
P.~Peixoto, S.~Ryu, A.~Bombrun, B.~Antonsson, K.~Kinnally, Mac inhibitors
  suppress mitochondrial apoptosis, Biochemical Journal 423 (2009) 381--387.

\bibitem{von_ahsen_2000}
O.~von Ahsen, C.~Renken, G.~Perkins, R.~M. Kluck, E.~Bossy-Wetzel, D.~D.
  Newmeyer, Preservation of mitochondrial structure and function after bid- or
  bax-mediated cytochrome c release, The Journal of Cell Biology 150 (2000)
  1027--1036.

\bibitem{rehm_single_cell_2002}
M.~Rehm, H.~Dussmann, R.~U. Janicke, J.~M. Tavare, D.~Kogel, J.~H.~M. Prehn,
  Single-cell fluorescence resonance energy transfer analysis demonstrates that
  caspase activation during apoptosis is a rapid process, The Journal of
  Biological Chemistry 277 (2002) 24506--24514.

\bibitem{madesh_2002}
M.~Madesh, B.~Antonsson, S.~M. Srinivasula, E.~S. Alnemri, G.~Hajnoczky, Rapid
  kinetics of tbid-induced cytochrome c and smac/diablo release and
  mitochondrial depolarization, The Journal of Biological Chemistry 277 (2002)
  5651--5659.

\bibitem{rehm_real_time_2003}
M.~Rehm, H.~Dussmann, J.~H.~M. Prehn, Real-time single cell analysis of
  smac/diablo release during apoptosis, The Journal of Cell Biology 162 (2003)
  1031--1042.

\bibitem{hellwig_2008}
C.~T. Hellwig, B.~F. Kohler, A.-K. Lehtivarjo, H.~Dussmann, M.~J. Courtney,
  J.~H.~M. Prehn, M.~Rehm, Real time analysis of tumor necrosis factor-related
  apoptosis-inducing ligand/cycloheximide-induced caspase activities during
  apoptosis initiation, The Journal of Biological Chemistry 283 (2008)
  21676--21685.

\bibitem{albeck_quantitative_2008}
J.~G. Albeck, J.~M. Burke, B.~B. Aldridge, M.~Zhang, D.~A. Lauffenburger, P.~K.
  Sorger, Quantitative analysis of pathways controlling extrinsic apoptosis in
  single cells, Molecular Cell 30 (2008) 11--25.

\bibitem{sun_2010}
T.~Sun, X.~Lin, Y.~Wei, Y.~Xu, P.~Shen, Evaluating bistability of bax
  activation switch, FEBS Letters 584 (2010) 954--960.

\bibitem{spencer_2011}
S.~L. Spencer, P.~K. Sorger, Measuring and modeling apoptosis in single cells,
  Cell 144 (2011) 926--939.

\bibitem{eissing_bistability_2004}
T.~Eissing, H.~Conzelmann, E.~D. Gilles, F.~Allgower, E.~Bullinger,
  P.~Scheurich, Bistability analyses of a caspase activation model for
  receptor-induced apoptosis, The Journal of Biological Chemistry 279 (2004)
  36892--36897.

\bibitem{legewie_mathematical_2006}
S.~Legewie, N.~Bluthgen, H.~Herzel, Mathematical modeling identifies inhibitors
  of apoptosis as mediators of positive feedback and bistability, PLoS
  Computational Biology 2 (2006) 1061--1073.

\bibitem{bagci_2006}
E.~Z. Bagci, Y.~Vodovotz, T.~R. Billiar, G.~B. Ermentrout, I.~Bahar,
  Bistability in apoptosis: Roles of bax, bcl-2, and mitochondrial permeability
  transition pores, Biophysical Journal 90 (2006) 1546--1559.

\bibitem{goldstein_2005}
J.~C. Goldstein, C.~Munoz-Pinedo, J.-E. Ricci, S.~R. Adams, A.~Kelekar,
  M.~Schuler, R.~Y. Tsien, D.~R. Green, Cytochrome c is released in a single
  step during apoptosis, Cell Death and Differentiation 12 (2005) 453--462.

\bibitem{cui_two_2008}
J.~Cui, C.~Chen, H.~Lu, T.~Sun, P.~Shen, Two independent positive feedbacks and
  bistability in the bcl-2 apoptotic switch, PLoS ONE 3 (2008) 1469.

\bibitem{chen_robustness_2007}
C.~Chen, J.~Cui, W.~Zhang, P.~Shen, Robustness analysis identifies the
  plausible model of the bcl-2 apoptotic switch, FEBS Letters 581 (2007)
  5143--5150.

\bibitem{harrington_2008}
H.~Harrington, K.~L. Lo, S.~Ghosh, K.~Tung, Construction and analysis of a
  modular model of caspase activation in apoptosis, Theoretical Biology and
  Medical Modelling 90 (2006) 1546--1559.

\bibitem{martinez-caballero_assembly_2009}
S.~Martinez-Caballero, L.~M. Dejean, M.~S. Kinnally, K.~J. Oh, C.~A. Mannella,
  K.~W. Kinnally, Assembly of the mitochondrial apoptosis-induced channel, mac,
  The Journal of Biological Chemistry 284 (2009) 12235--12245.

\bibitem{leber_embedded_2007}
B.~Leber, J.~Lin, D.~W. Andrews, Embedded together: The life and death
  consequences of interaction of the bcl-2 family with membranes, Apoptosis: an
  international journal on programmed cell death 12 (2007) 897--911.

\bibitem{brunelle_control_2009}
J.~K. Brunelle, A.~Letai, Control of mitochondrial apoptosis by the bcl-2
  family, Journal of Cell Science 122 (2009) 437--441.

\bibitem{chipuk_do_2008}
J.~E. Chipuk, D.~R. Green, How do bcl-2 proteins induce mitochondrial outer
  membrane permeabilization?, Trends in Cell Biology 18 (2008) 157--164.

\bibitem{luo_bidbcl2_1998}
X.~Luo, I.~Budihardjo, H.~Zou, C.~Slaughter, X.~Wang, Bid, a bcl2 interacting
  protein, mediates cytochrome c release from mitochondria in response to
  activation of cell surface death receptors, Cell 94 (1998) 481--490.

\bibitem{desagher_bid-induced_1999}
S.~Desagher, A.~Osen-Sand, A.~Nichols, R.~Eskes, S.~Montessuit, S.~Lauper,
  K.~Maundrell, B.~Antonsson, J.~C. Martinou, Bid-induced conformational change
  of bax is responsible for mitochondrial cytochrome c release during
  apoptosis, The Journal of cell biology 144 (1999) 891--901.

\bibitem{ott_mitochondria_2007}
M.~Ott, V.~Gogvadze, S.~Orrenius, B.~Zhivotovsky, Mitochondria, oxidative
  stress and cell death, Apoptosis 12 (2007) 913--922.

\bibitem{willis_apoptosis_2007}
S.~N. Willis, J.~I. Fletcher, T.~Kaufmann, M.~F. van Delft, L.~Chen, P.~E.
  Czabotar, H.~Ierino, E.~F. Lee, W.~D. Fairlie, P.~Bouillet, Apoptosis
  initiated when bh3 ligands engage multiple bcl-2 homologs, not bax or bak,
  Science 315 (2007) 856--859.

\bibitem{cheng_bcl-2_2001}
E.~H. Cheng, M.~C. Wei, S.~Weiler, R.~A. Flavell, T.~W. Mak, T.~Lindsten, S.~J.
  Korsmeyer, Bcl-2, bcl-xl sequester bh3 domain-only molecules preventing
  bax-and bak-mediated mitochondrial apoptosis, Molecular Cell 8 (2001)
  705--711.

\bibitem{kim_hierarchical_2006}
H.~Kim, M.~Rafiuddin-Shah, H.~Tu, J.~R. Jeffers, G.~P. Zambetti, J.~J.~D.
  Hsieh, E.~H.~Y. Cheng, Hierarchical regulation of mitochondrion-dependent
  apoptosis by bcl-2 subfamilies, Nature Cell Biology 8 (2006) 1348--1358.

\bibitem{willis_proapoptotic_2005}
S.~N. Willis, L.~Chen, G.~Dewson, A.~Wei, E.~Naik, J.~I. Fletcher, J.~M. Adams,
  D.~C.~S. Huang, Proapoptotic bak is sequestered by mcl-1 and bcl-xl, but not
  bcl-2, until displaced by bh3-only proteins, Genes \& Development 19 (2005)
  1294--1305.

\bibitem{kuwana_bid_2002}
T.~Kuwana, M.~R. Mackey, G.~Perkins, M.~H. Ellisman, M.~Latterich,
  R.~Schneiter, D.~R. Green, D.~D. Newmeyer, Bid, bax, and lipids cooperate to
  form supramolecular openings in the outer mitochondrial membrane, Cell 111
  (2002) 331--342.

\bibitem{terrones_lipidic_2004}
O.~Terrones, B.~Antonsson, H.~Yamaguchi, H.~G. Wang, J.~Liu, R.~M. Lee,
  A.~Hermann, G.~Basanez, Lipidic pore formation by the concerted action of
  proapoptotic bax and tbid, Journal of Biological Chemistry 279 (2004)
  30081--30091.

\bibitem{kuwana_bh3_2005}
T.~Kuwana, L.~Bouchier-Hayes, J.~E. Chipuk, C.~Bonzon, B.~A. Sullivan, D.~R.
  Green, D.~D. Newmeyer, Bh3 domains of bh3-only proteins differentially
  regulate bax-mediated mitochondrial membrane permeabilization both directly
  and indirectly, Molecular Cell 17 (2005) 525--535.

\bibitem{antonsson_2001}
B.~Antonsson, S.~Montessuit, B.~Sanchez, J.-C. Martinou, Bax is present as a
  high molecular weight oligomer/complex in the mitochondrial membrane of
  apoptotic cells, The Journal of Biological Chemistry 276 (2001) 11615--11623.

\bibitem{valentijn_2008}
A.~Valentijn, J.-P. Upton, P.~Gilmore, Analysis of endogenous bax complexes
  during apoptosis using blue native page: implications for bax activation and
  oligomerization, Biochemical Journal 412 (2008) 347--357.

\bibitem{hua_effects_2005}
F.~Hua, M.~Cornejo, M.~H. Cardone, C.~L. Stokes, D.~A. Lauffenburger, Effects
  of bcl-2 levels on fas signaling-induced caspase-3 activation: molecular
  genetic tests of computational model predictions, Journal of Ifmmunology 175
  (2005) 985--995.

\bibitem{dlugosz_bcl-2_2006}
P.~J. Dlugosz, L.~P. Billen, M.~G. Annis, W.~Zhu, Z.~Zhang, J.~Lin, B.~Leber,
  D.~W. Andrews, Bcl-2 changes conformation to inhibit bax oligomerization, The
  EMBO Journal 25 (2006) 2287--2296.

\bibitem{letai_distinct_2002}
A.~Letai, M.~C. Bassik, L.~D. Walensky, M.~D. Sorcinelli, S.~Weiler, S.~J.
  Korsmeyer, Distinct bh3 domains either sensitize or activate mitochondrial
  apoptosis, serving as prototype cancer therapeutics, Cancer Cell 2 (2002)
  183--192.

\bibitem{walensky_activation_2004}
L.~D. Walensky, A.~L. Kung, I.~Escher, T.~J. Malia, S.~Barbuto, R.~D. Wright,
  G.~Wagner, G.~L. Verdine, S.~J. Korsmeyer, Activation of apoptosis in vivo by
  a hydrocarbon-stapled bh3 helix, American Association for the Advancement of
  Science 305 (2004) 1466--1470.

\bibitem{chen_modeling_2007}
C.~Chen, J.~Cui, H.~Lu, R.~Wang, S.~Zhang, P.~Shen, Modeling of the role of a
  bax-activation switch in the mitochondrial apoptosis decision, Biophysical
  Journal 92 (2007) 4304--4315.

\bibitem{ramsey_dizzy:_2005}
S.~Ramsey, D.~Orrell, H.~Bolouri, Dizzy: stochastic simulation of large-scale
  genetic regulatory networks, Journal of Bioinformatics and Computational
  Biology 3 (2005) 415--436.

\bibitem{hucka_2003}
M.~Hucka, A.~Finney, H.~M. Sauro, H.~Bolouri, J.~C. Doyle, H.~Kitano, The
  systems biology markup language (sbml): a medium for representation and
  exchange of biochemical network models, Bioinformatics 19 (2003) 524--531.

\bibitem{olivier_2005}
O.~B. G, J.~M. Rohwer, J.-H. Hofmeyr, Modelling cellular systems with pysces,
  Bioinformatics 21 (2005) 560--561.

\bibitem{eissing_steady_2007}
T.~Eissing, S.~Waldherr, F.~Allgower, P.~Scheurich, E.~Bullinger, Steady state
  and bi-stability evaluation of simple protease signalling networks,
  Biosystems 90 (2007) 591--601.

\bibitem{angeli_multi-stability_2003}
D.~Angeli, E.~D. Sontag, Multi-stability in monotone input/output systems,
  System \& Control Letters 51 (2003) 185--202.

\bibitem{tyson_sniffers_2003}
J.~J. Tyson, K.~C. Chen, B.~Novak, Sniffers, buzzers, toggles and blinkers:
  dynamics of regulatory and signaling pathways in the cell, Current Opinion in
  Cell Biology 15 (2003) 221--231.

\bibitem{ferrell_trippingswitch_1996}
J.~E.~J. Ferrell, Tripping the switch fantastic: how a protein kinase cascade
  can convert graded inputs into switch-like outputs, Trends in biochemical
  sciences 21 (1996) 460--466.

\bibitem{legewie_quantitative_2005}
S.~Legewie, N.~Bluthgen, H.~Herzel, Quantitative analysis of ultrasensitive
  responses, FEBS Journal 272 (2005) 4071--4079.

\bibitem{ruffolo_bcl-2_2003}
S.~C. Ruffolo, G.~C. Shore, Bcl-2 selectively interacts with the bid-induced
  open conformer of bak, inhibiting bak auto-oligomerization, The Journal of
  Biological Chemistry 278 (2003) 25039--25045.

\bibitem{tan_auto-activation_2006}
C.~Tan, P.~J. Dlugosz, J.~Peng, Z.~Zhang, S.~M. Lapolla, S.~M. Plafker, D.~W.
  Andrews, J.~Lin, Auto-activation of the apoptosis protein bax increases
  mitochondrial membrane permeability and is inhibited by bcl-2, Journal of
  Biological Chemistry 281 (2006) 14764--14775.

\bibitem{dussmann_single-cell_2009}
H.~Dussmann, M.~Rehm, C.~G. Concannon, S.~Anguissola, M.~Wurstle, S.~Kacmar,
  P.~Voller, H.~J. Huber, J.~H.~M. Prehn, Single-cell quantification of bax
  activation and mathematical modelling suggest pore formation on minimal
  mitochondrial bax accumulation, Cell Death and Differentiation 17 (2009)
  278--290.

\bibitem{ferrell_responses_1997}
J.~E.~J. Ferrell, How responses get more switch-like as you move down a protein
  kinase cascade, Trends in Biochemical Sciences 22 (1997) 288--289.

\bibitem{ferrell_regulated_1998}
J.~E.~J. Ferrell, How regulated protein translocation can produce switch-like
  responses, Trends in Biochemical Sciences 23 (1998) 461--465.

\bibitem{angeli_detection_2004}
D.~Angeli, J.~E. Ferrell, E.~D. Sontag, Detection of multistability,
  bifurcations, and hysteresis in a large class of biological positive-feedback
  systems, Proceedings of the National Academy of Sciences 101 (2004)
  1822--1827.

\bibitem{kitano_systems_2002}
H.~Kitano, Systems biology: a brief overview, Science 295 (2002) 1662--1664.

\bibitem{bluthgen_robust_2003}
N.~Bluthgen, H.~Herzel, How robust are switches in intracellular signaling
  cascades?, Journal of Theoretical Biology 225 (2003) 293--300.

\bibitem{barkai_robustness_1997}
N.~Barkai, S.~Leiber, Robustness in simple biochemical networks, Nature 387
  (1997) 913--917.

\bibitem{fell_metabolic_1992}
D.~A. Fell, Metabolic control analysis: a survey of its theoretical and
  experimental development, Biochemical Journal 286 (1992) 313--330.

\bibitem{kholodenko_quantification_1997}
B.~N. Kholodenko, J.~B. Hoek, H.~V. Westerhoff, G.~C. Brown, Quantification of
  information transfer via cellular signal transduction pathways, FEBS letters
  414 (1997) 430--434.

\end{thebibliography}
\bibliographystyle{elsarticle-num}

\section*{Figures}

\begin{figure}[H]
\centering
\includegraphics{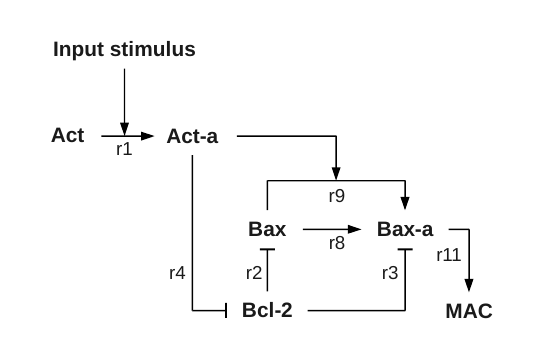}
\caption{Simplified scheme of the hybrid model used in this work (production and degradation flows, MAC growth and reverse reactions are not depicted). Reactions r1-r11 are numbered in accordance with the model description in section~\ref{sec:biol_relevance}.}
\end{figure} 

\begin{figure}[H]
\centering
\includegraphics{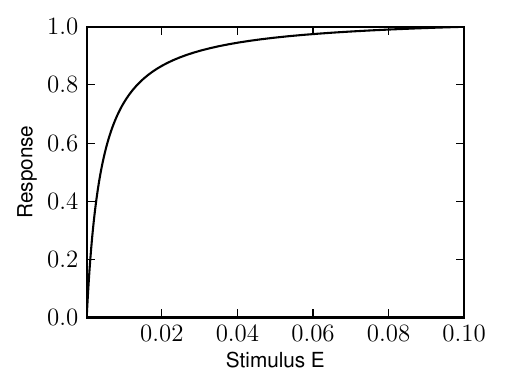}
\caption{The stimulus-response dependence of the hybrid model using reference parameters listed in the Table 3. The response is normalized to one.}
\end{figure}

\begin{figure}[H]
\centering
\includegraphics{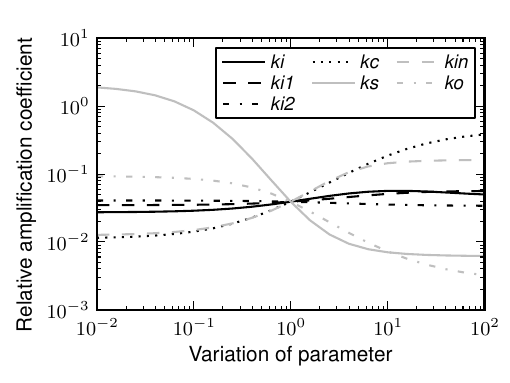}
\caption{Relative amplification coefficients of the hybrid model. On x-axis - the variation of single reaction parameter (in inset) from the reference values (Table 3).}
\end{figure}

\begin{figure}[H]
\centering
\includegraphics{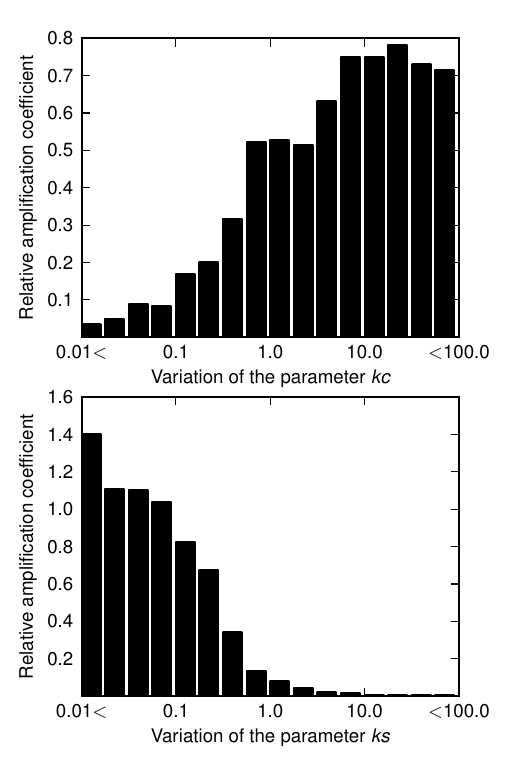}
\caption{The distribution of the mean relative amplification coefficient for the reaction parameters \emph{kc} and \emph{ks}. The averaging is done over the remaining parameters chosen at random}
\end{figure}

\begin{figure}[H]
\centering
\includegraphics{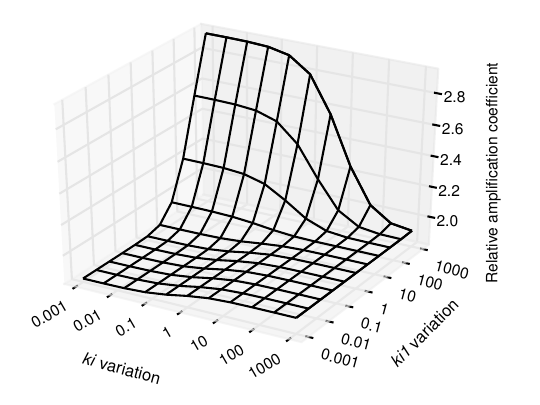}
\caption{Dependence of relative amplification coefficient on simultaneous variation of reaction parameters \emph{ki} and \emph{ki1}. Parameters \emph{ki}/\emph{ki1} are the reaction rates of binding and inhibition of inactive/active effectors by their anti-apoptotic relatives.}
\end{figure}

\begin{figure}[H]
\centering
\includegraphics{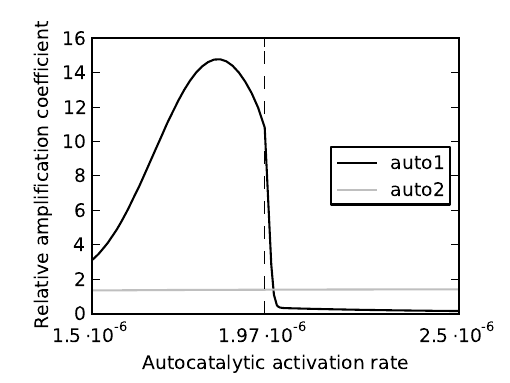}
\caption{Dependence of relative amplification coefficient on the variation of reaction rate of Bax auto-activation. Black line corresponds to the auto-activation mediated by monomers (as described by reaction auto1), gray line corresponds to the auto-activation mediated by oligomerized Bax (reaction auto2).}
\end{figure}

\begin{figure}[H]
\centering
\includegraphics{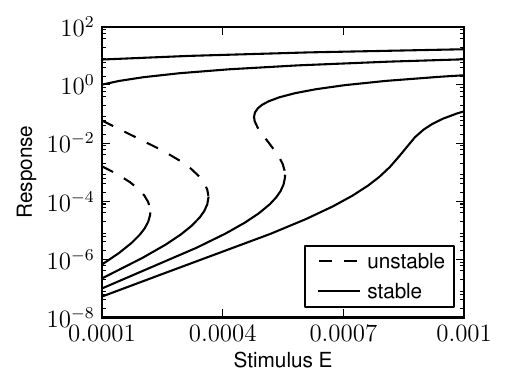}
\caption{ Steady-state stimulus-response dependencies of the HM with addition of the auto-activation (auto1). Curves, from right to left correspond to \emph{ka} values $1.5\!\times\!10^{-6}$, $1.6\!\times\!10^{-6}$, $1.7\!\times\!10^{-6}$ and $1.8\!\times\!10^{-6}$. 
As \emph{ka} value exceeds certain threshold, bistability emerges. Further strengthening of the auto-activation expands the bistable range and shifts this range to the left along the stimulus axis. Overly fast auto-activation ($ka > 1.97\!\times\!10^{-6}$) makes model's response to basal stimulation ($1.0\!\times\!10^{-4}$) inadequately high.}
\end{figure}

\begin{figure*}[H]
\centering
\includegraphics[width=1.0\textwidth]{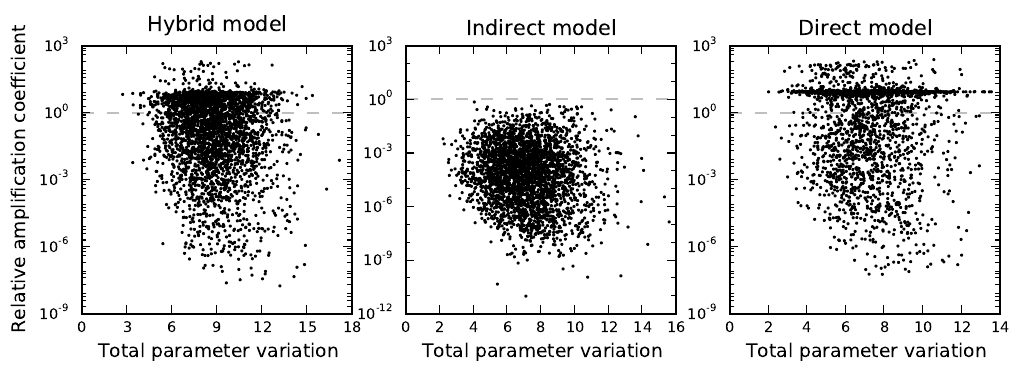}
\caption{Scatter plots of relative amplification coefficient as a function of total parameter variation. Dots localized over gray dashed line correspond to randomly generated parameter sets that yield ultrasensitive response, i.e. those with relative amplification coefficient higher than one.}
\end{figure*}


\begin{figure*}[H]
\centering
\includegraphics[width=1.0\textwidth]{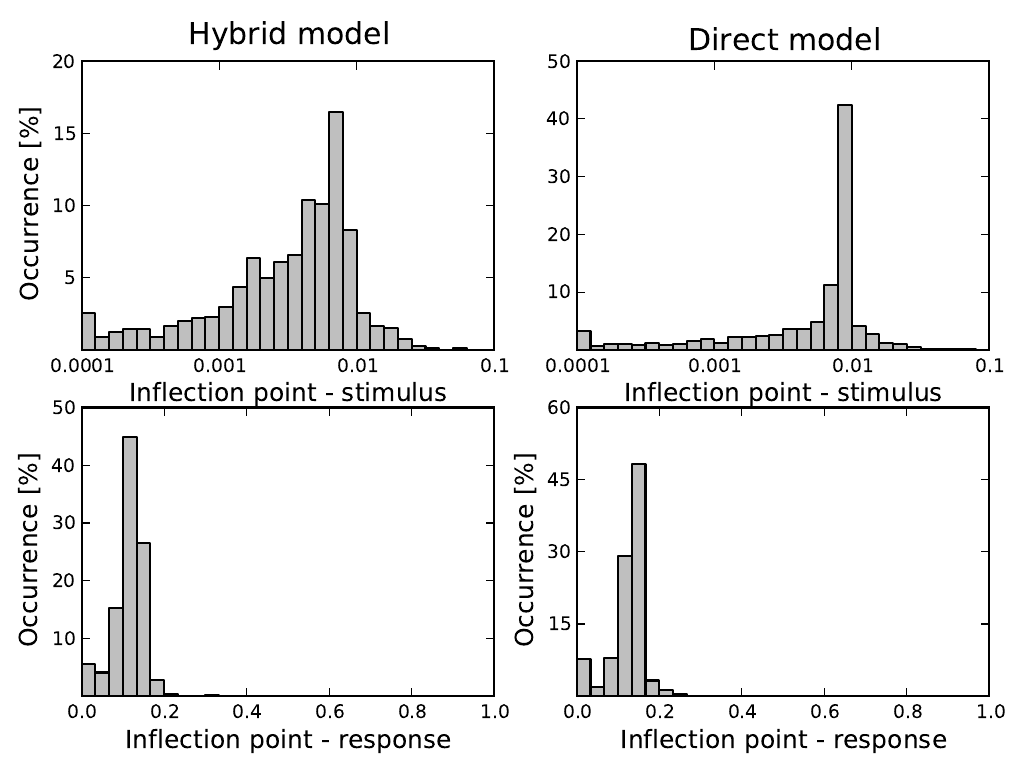}
\caption{For each stimulus-response dependence yielding an ultrasensitive response (i.e. the sets represented by dots placed above gray dashed line in Fig. 8) we have measured its threshold (inflection point at the stimulus response curve) stimulus and corresponding threshold response. Histograms show distribution of these values.}
\end{figure*}

\section*{Tables}

\begin{table*}[H]\small
\begin{center}
\makebox[\textwidth]{
\begin{tabular}{@{}lrrr@{}} \toprule
Species	&	\multicolumn{2}{c}{Initial concentration}	& Notes \& Ref.	\\
&	$\# \cdot cell^{-1}$		& $nM$				\\ \midrule
Bcl-2	&	60 000		& 100	&~\cite{kuwana_bcl-2-family_2003,chen_robustness_2007}	\\
Bax	&	120 000		& 200	&~\cite{kuwana_bcl-2-family_2003,chen_robustness_2007}	\\
Act	&	6000		& 10	&~\cite{hua_effects_2005,dlugosz_bcl-2_2006}		\\
Bax--a	&	0		& 0	\\
Act--a	&	0		& 0	\\
Bcl-2$\sim$Bax	&	0	& 0	\\
Bcl-2$\sim$Bax--a	&	0	& 0	\\
Bcl-2$\sim$Act--a	&	0	& 0	\\
MACi			&	0	& 0	\\ \bottomrule
\end{tabular}
}
\end{center}
\caption{List of initial concentrations. Concentrations are listed as the absolute number of molecules per reference cell volume 1 pl as well as in more common units - nanomols.}
\end{table*}

\begin{table*}[H]\small
\begin{center}
\makebox[\textwidth]{
\begin{tabular}{@{}llll@{}} \toprule
No.	& Reaction 			& $k_{+}$	& $k_{-}$	\\ \midrule
1. 	& Act $\xrightarrow{E}$ Act--a	& \emph{E}	&   		\\
2. 	& Bcl-2 + Bax $\rightleftharpoons$ Bcl-2$\sim$Bax	& \emph{ki}	& \emph{kmi} \\
3. 	& Bcl-2 + Bax--a $\rightleftharpoons$ Bcl-2$\sim$Bax--a	& \emph{ki1}	& \emph{kmi1} \\
4.	& Bcl-2 + Act--a $\rightleftharpoons$ Bcl-2$\sim$Act--a & \emph{ki2}	& \emph{kmi2} \\ 
5.	& Bcl-2$\sim$Act--a + Bax $\rightleftharpoons$ Bcl-2$\sim$Bax + Act--a	& \emph{ki}	& \emph{ki2}	\\
6.	& Bcl-2$\sim$Act--a + Bax--a $\rightleftharpoons$ Bcl-2$\sim$Bax--a + Act--a	& \emph{ki1}	& \emph{ki2} \\
7.	& Bcl-2$\sim$Bax--a + Bax $\rightleftharpoons$ Bcl-2$\sim$Bax + Bax--a		& \emph{ki}	& \emph{ki1} \\
8.	& Bax $\rightarrow$ Bax--a	& \emph{ks}	&	\\
9.	& Act--a + Bax $\rightarrow$ Act--a + Bax--a	& \emph{kc}		&  \\
10.	& Bax--a $\rightarrow$ Bax			& \emph{kin}		&  \\
11.	& Bax--a + Bax--a $\rightleftharpoons$ MAC$_{2}$	& \emph{ko}	& \emph{kmo}  \\
12.	& Bax--a + MAC$_{i}$ $\rightleftharpoons$ MAC$_{i+1}$	& \emph{ko}	& \emph{kmo} \\
13.	& (All) $\rightarrow$				& \emph{kd}		&  \\
14.	& $\rightarrow$ Bcl-2				& \emph{kpBcl-2}	&  \\
15.	& $\rightarrow$ Bax				& \emph{kpBax}		&  \\
16.	& $\rightarrow$ Act				& \emph{kpAct}		&  \\ \bottomrule
\end{tabular}
}
\end{center}
\caption{List of reactions and corresponding reaction parameters. Reversible reactions are listed with forward ($k_{+}$ column) and reverse ($k_{-}$ column) reaction parameters in the same row.}
\end{table*}

\begin{table*}[H]\small
\begin{center}
\makebox[\textwidth]{
\begin{tabular}{@{}lrlrll@{}} \toprule
Param.	& \multicolumn{2}{c}{Value} & \multicolumn{2}{c}{Unit} & Notes \& Ref. \\ \midrule
\emph{ki} 	& $1.0\!\times\!10^{-6}$ & $(1.0\!\times\!10^{4})$ & $cell\!\cdot\!\#^{-1}\!\cdot\!min^{-1}$ & $(M^{-1}\!\cdot\!s^{-1})$ & (in direct) $= 0$, estimated	\\
\emph{kmi}	& $0.06$ &$(0.001)$ & $min^{-1}$ & $(s^{-1})$	& = \emph{kmi1}	\\
\emph{ki1}	& $1.0\!\times\!10^{-6}$ & $(1.0\!\times\!10^{4})$ & $cell\!\cdot\!\#^{-1}\!\cdot\!min^{-1}$ & $(M^{-1}\!\cdot\!s^{-1})$ & (in indirect) $= 0$, estimated	\\
\emph{kmi1}	& $0.06$ &$(0.001)$ & $min^{-1}$ & $(s^{-1})$	& same as in ~\cite{cui_two_2008}	\\
\emph{ki2}	& $1.0\!\times\!10^{-6}$ & $(1.0\!\times\!10^{4})$ & $cell\!\cdot\!\#^{-1}\!\cdot\!min^{-1}$ & $(M^{-1}\!\cdot\!s^{-1})$ & $K_d \doteq 100nM$~\cite{letai_distinct_2002,walensky_activation_2004} \\
\emph{kmi2}	& $0.06$ &$(0.001)$ & $min^{-1}$ & $(s^{-1})$	& same as in ~\cite{cui_two_2008}	\\
\emph{kc}	& $1.0\!\times\!10^{-6}$ & $(1.0\!\times\!10^{4})$ & $cell\!\cdot\!\#^{-1}\!\cdot\!min^{-1}$ & $(M^{-1}\!\cdot\!s^{-1})$ & (in indirect) $= 0$, estimated \\
\emph{ks}	& $0.06$ &$(0.001)$ & $min^{-1}$ & $(s^{-1})$	& estimated	\\
\emph{kin}	& $0.06$ &$(0.001)$ & $min^{-1}$ & $(s^{-1})$	& estimated	\\
\emph{ko}    & $1.0\!\times\!10^{-6}$ & $(1.0\!\times\!10^{4})$ & $cell\!\cdot\!\#^{-1}\!\cdot\!min^{-1}$ & $(M^{-1}\!\cdot\!s^{-1})$ & estimated	\\
\emph{kmo}	& $0.06$ &$(0.001)$ & $min^{-1}$ & $(s^{-1})$	& estimated	\\
\emph{kd}	& $0.0039$ & $(6.5\!\times\!10^{-5})$ & $min^{-1}$ & $(s^{-1})$	& $t_{1/2}\doteq180\;min$, estimated \\
\emph{kpBcl-2}	& $234.0$ & $(6.4\!\times\!10^{-12})$ & $\#\!\cdot\!cell^{-1}\!\cdot\!min^{-1}$ & $(M\!\cdot\!s^{-1})$ & $=$ Bcl-2$_{init}\cdot$\emph{kd}\\
\emph{kpBax}	& $468.0$ & $(1.3\!\times\!10^{-11})$ & $\#\!\cdot\!cell^{-1}\!\cdot\!min^{-1}$ & $(M\!\cdot\!s^{-1})$ & $=$ Bax$_{init}\cdot$\emph{kd}\\
\emph{kpAct}	& $23.4$ & $(6.4\!\times\!10^{-13})$ & $\#\!\cdot\!cell^{-1}\!\cdot\!min^{-1}$ & $(M\!\cdot\!s^{-1})$ & $=$ Act$_{init}\cdot$\emph{kd}\\ \bottomrule
\end{tabular}
}
\end{center}
\caption{List of reference values of reaction parameters. By setting parameters \emph{ki1} and \emph{kc} equal to zero, the hybrid model is reduced to the indirect one. Similarly, the direct model can be obtained by setting parameters \emph{ki} and \emph{ks} of the hybrid model equal to zero.  Production rates \emph{kpBcl-2}, \emph{kpBax} and \emph{kpAct} have been set to balance degradation of corresponding species under initial conditions.}
\end{table*}

\begin{table*}[H]\small
\begin{center}
\makebox[\textwidth]{
\begin{tabular}{@{}lrlrl@{}} \toprule
Param.	& \multicolumn{2}{c}{Value} & \multicolumn{2}{c}{Unit} \\ \midrule
\emph{ki} 	& $1.0\!\times\!10^{-8}$ & $(1.0\!\times\!10^{2})$ & $cell\!\cdot\!\#^{-1}\!\cdot\!min^{-1}$ & $(M^{-1}\!\cdot\!s^{-1})$ \\
\emph{ki1}	& $1.0\!\times\!10^{-4}$ & $(1.0\!\times\!10^{6})$ & $cell\!\cdot\!\#^{-1}\!\cdot\!min^{-1}$ & $(M^{-1}\!\cdot\!s^{-1})$ \\
\emph{kc}	& $2.0\!\times\!10^{-5}$ & $(2.0\!\times\!10^{5})$ & $cell\!\cdot\!\#^{-1}\!\cdot\!min^{-1}$ & $(M^{-1}\!\cdot\!s^{-1})$ \\
\emph{ka}	& $1.8\!\times\!10^{-6}$ & $(1.8\!\times\!10^{4})$ & $cell\!\cdot\!\#^{-1}\!\cdot\!min^{-1}$ & $(M^{-1}\!\cdot\!s^{-1})$ \\
\emph{ks}	& $6.0\!\times\!10^{-4}$ & $(1.0\!\times\!10^{-5})$ & $min^{-1}$ & $(s^{-1})$	\\ \bottomrule
\end{tabular}
}
\end{center}
\caption{List of adjusted values of reaction parameters. In order to obtain desirable level of ultrasensitivity, the listed parameters have been modified according to results of previous simulations. The other parameters listed in the Table 3 remained unchanged.}
\end{table*}

\end{document}